\documentstyle[11pt]{article}

\newcommand{\sect}[1]{\setcounter{equation}{0}\section{#1}\indent}
\renewcommand{\theequation}{\thesection.\arabic{equation}}
\def\appendix#1{\addtocounter{section}{1}\setcounter{equation}{0}
\renewcommand{\thesection}{\Alph{section}}
\renewcommand{\theequation}{\Alph{section}.\arabic{equation}}
\section*{Appendix \thesection\protect\indent #1}
\addcontentsline{toc}{section}{Appendix \thesection #1}}
\def\fnote#1#2{\begingroup\def\thefootnote{#1}\footnote{#2}\addtocounter
{footnote}{-1}\endgroup}

\font\mathbold=cmmib10 at12pt
\font\bf=cmbx10 at12pt
\newfam\bffam 
\textfont\bffam=\mathbold
\mathchardef\beta="710C
\mathchardef\gamma="710D
\mathchardef\eta="7111
\mathchardef\xi="7118
\newcommand{\BE}{\begin{equation}}
\newcommand{\EE}{\end{equation}}
\newcommand{\bea}{\begin{eqnarray}}
\newcommand{\eea}{\end{eqnarray}}
\newcommand{\ba}{\begin{array}}
\newcommand{\ea}{\end{array}}
\newcommand{\vs}[1]{\vspace{#1 mm}}
\newcommand{\hs}[1]{\hspace{#1 mm}}
\newcommand{\nonum}{\nonumber}
\def\[{{[}}
\def\]{{]}}

\def\L#1#2{{L_{#1}}^{#2}} 
\def\Lh#1#2{{{\hat L}_{#1}}^{#2}} 
\def\Lt#1#2{{{\tilde L}_{#1}}^{#2}} 
\def\Lth#1#2{{ {\hat {\tilde L} }_{#1}}^{#2}}
\def\R#1#2{{R_{#1}}^{#2}} 
 
\def\to#1{\stackrel{#1}{\longrightarrow}}

\renewcommand{\a}{\alpha}
\renewcommand{\b}{\beta}
\newcommand{\de}{\delta}
\newcommand{\ep}{\epsilon}

\def\ga{\gamma}

\def\na{\natural}

\def\sh{\sharp}

\def\PhiI{\Phi\hs{-1.4}{\rm I}}
\def\PsiI{\Psi\hs{-1.7}{\rm I}}

\newcommand{\half}{\frac{1}{2}}

\def\CM{{\cal M}}

\def\CF{{\cal F}}

\def\CW{{\cal W}}

\def\cU{{\cal U}}

\def\Qt{\tilde Q}

\def\S{\Sigma}

\def\thetab{{\bar \theta}}

\def\bh{{\hat b}}
\def\Phih{{\hat \Phi}}

\def\phit{{\tilde \phi}}

\def\thetat{{\tilde \theta}}
\def\thetatb{{\bar \thetat}}

\def\zetat{{\tilde \zeta}}

\def\St{{\tilde \S}}
\def\Zt{{\tilde Z}}
\def\Qt{{\tilde Q}}

\newcommand{\NP}[1]{Nucl.\ Phys.\ {\bf #1}}
\newcommand{\PL}[1]{Phys.\ Lett.\ {\bf #1}}

\newcommand{\PR}[1]{Phys.\ Rev.\ {\bf #1}}

\begin{document}
\topmargin -35pt
\oddsidemargin 0mm
\begin{titlepage}
\setcounter{page}{0}
\vs{15}
\begin{flushright}
KEK Preprint 98-200\\
KEK-TH-600\\
JHEP02(1999)017\\
\end{flushright}
\vs{0}
\begin{center}
\vs{15}
{\bf {\Large $(p,q)$-strings and New Spacetime Superalgebras
}}\\
\vs{40}

{\large Makoto SAKAGUCHI\fnote{\star}{e-mail:
sakagu@post.kek.jp
}}\\
\vs{7}
{\em
Theory Division, Institute of Particle and Nuclear Studies,\\
High Energy Accelerator Research Organization (KEK),\\
1-1 Oho, Tsukuba, Ibaraki, 305-0801
Japan}\\
\end{center}
\vs{30}

\baselineskip 16pt
\centerline{\bf Abstract}
\vs{5}

We present a geometric formulation
of $(p,q)$-strings in which
the $Sl(2;Z)$-doublet of the two-form gauge potentials
is constructed as second order in the supersymmetric currents.
The currents are constructed
using a supergroup manifold corresponding to
the $(p,q)$-string superalgebra,
which contains fermionic generators in addition to the supercharges
and transforms under the $Sl(2;Z)$.
The properties of the superalgebra
and the generalizations to higher $p$-branes
are discussed.
\end{titlepage}
\newpage
\baselineskip 18pt
\newpage
\renewcommand{\thefootnote}{\arabic{footnote}}
\setcounter{footnote}{0}
\sect{Introduction}
It is now widely appreciated that
the IIB superstring theory enjoys $Sl(2;Z)$-symmetry
non-perturbatively.
$(p,q)$-strings~\cite{pqstrings} has been
studied~\cite{Townsend1,Townsend2,pqstrings2}
as a formulation
in which the $Sl(2;Z)$-symmetry is manifest.
On the other hand,
a manifestly supersymmetric formulation of the Green-Schwarz superstring,
based on a superalgebra
which was obtained as the global limit~\cite{Green}
of a superalgebra found by Siegel~\cite{Siegel1},
was given by Siegel~\cite{Siegel}.
The superalgebra is a generalization of super Poincar\'e algebra,
and contains a new fermionic generator.
Constructing a suitable set of supercurrents on the corresponding
supergroup manifold,
he wrote down the Wess-Zumino term of the
Green-Schwarz superstring in a second-order expression of
the supercurrents.
Using this formulation,
$p$-branes ($p>1$), found in the brane scan~\cite{scan}, were
formulated by Bergshoeff and Sezgin~\cite{BS}.
Introducing a set of superalgebras,
they wrote the Wess-Zumino terms
for the $p$-branes,
which are $(p+1)$-th order in the supercurrents.
These $p$-brane ($p\geq 1$) minimal
superalgebras were further investigated
in \cite{Alex}.
Following the same line,
non-minimal 
IIA strings and IIB F- and D-strings were formulated in \cite{sakagu},
using the IIA superalgebra and the F- and  D-string superalgebra,
respectively.

In this paper, we consider $(p,q)$-strings
using this formulation.
It is known~\cite{sakagu}
that the pullback to the F(D)-string worldsheet of
the NS$\otimes$NS(R$\otimes$R)
gauge potential can be
obtained from the F(D)-string superalgebra.
In order to construct $(p,q)$-strings,
we need the pullback to the same worldsheet
of the NS$\otimes$NS and the R$\otimes$R gauge potentials.
We obtain the pullback to the F-string worldsheet
of the R$\otimes$R gauge potential
(or, the pullback to the D-string worldsheet
of the NS$\otimes$NS gauge potential), in sec.3,
in the following two ways: 
First, we consider a map which transforms the D-string superalgebra
to the F-string superalgebra.
Using this map, we obtain the pullback to the F-string worldsheet
of the R$\otimes$R
gauge potential
explicitly.
Second,
we construct the pullback to the F-string worldsheet
of the R$\otimes$R gauge potential,
starting from a superalgebra
which is obtained by interchanging fermionic generators,
$\S^\a$ and $\St^\a$, in the D-string superalgebra.
This is a nontrivial feature of our formulation
which was suggested in~\cite{sakagu}:
Interchanging fermionic generators
results in interchanging the F- and the D-string worldsheets
(or, exchanging the R$\otimes$R two-form gauge potential for
the NS$\otimes$NS one),
In order to confirm this feature,
we show that the pullback to the F-string worldsheet of the R$\otimes$R
gauge potential can be obtained from
a superalgebra which is obtained by interchanging fermionic
generators in the F-string superalgebra.
After a quick review of the explicit construction of
the $Sl(2;Z)$-doublet of two-form gauge potentials
in terms of $b_R$ and $b_{NS}$,
we present the $(p,q)$-string superalgebra,
and show that the $Sl(2;Z)$-doublet of two-form gauge potentials
can be constructed by using the superalgebra, in sec. 4.
It is observed that
at a special point on the $Sl(2;Z)$-orbit, $\chi=\phi=0$,
the $(p,q)$-string superalgebra
reduces to the IIB superalgebra
which is T-dual
to the IIA superalgebra in presence of
F-string, D0- and D2-branes.
The last section is devoted to a summary and discussions
of generalizations to the D3-brane, the $(p,q)$ 5-brane
and the KK5-brane.
\sect{F- and D-strings}
We first review the formulation of IIB F- and D-strings
following~\cite{sakagu}.
The F-string superalgebra is generated by
supertranslations and F-string charges
$T_A=\{Q_A, Z^A\}$, where $Q_A=(P_a, Q_\a,\Qt_\a)$
and $Z^A=(Z^a, Z^\a, \tilde Z^\a)$,
as
\bea
\{Q_{\a I},Q_{\b J}\} &=&
    (\ga^a\otimes I_{IJ})_{\a\b}P_a
    -(\ga_a\otimes {\sigma_3}_{IJ})_{\a\b}Z^a,\nonum\\
\[P_a,Q_{\b I}\]&=&-2(\ga_a\otimes {\sigma_3}_{IJ})_{\a\b}Z^{\a J},
\label{Falgebra}\\
\[Z^a,Q_{\b I}\]&=&2(\ga^a\otimes I_{IJ})_{\a\b}Z^{\a J},
\nonum
\eea
where we employed a compact notation,
$Q_{\a I}=\left( \ba{@{\,}c@{\,}} \Qt_\a \\ Q_\a \ea \right) $, e.t.c.
The indices $I$ and $J$ are frequently omitted below.
The left-invariant supergroup vielbeins $\L{M}{A}$
and the pullback one-form $\L{}{A}$
are obtained by the left-invariant Maurer-Cartan one-form,
\bea
U^{-1}dU = dZ^M\L{M}{A}T_A=\L{}{A}T_A,
\eea
where $Z^M$ are coordinates on the supergroup manifold,
and $U$ is a supergroup element.
Parametrizing the supergroup manifold by
\bea
U=e^{z_aZ^a}e^{\zetat_\a \Zt^\a}e^{\zeta_\a Z^\a}
e^{x^aP_a}e^{\thetat^\a \Qt_\a}e^{\theta^\a Q_\a},
\eea
the pullback left-invariant vielbeins
of the supergroup manifold, $L^A$,
were obtained
in terms of coordinates on the supergroup manifold.
Similarly, the right-invariant supergroup vielbeins $\R{M}{A}$
are obtained
by the right-invariant Maurer-Cartan one-form,
\bea
dUU^{-1}=dZ^M\R{M}{A}T_A.
\eea
The rigid supersymmetry transformations are
interpreted as being the part
of an infinitesimal transformation, $\de Z^M=\ep^A\R{A}{M}$,
characterized by $\ep^\a$,
where $\R{A}{M}$ is the inverse of $\R{M}{A}$.
The pullback left-invariant vielbeins, $L^A$,
are invariant under the supersymmetry transformation,
by definition.

The pullback to the F-string worldsheet of the NS$\otimes$NS two-form
gauge potential
was defined as a second-order expression in supercurrents,
\bea
B_{NS}=\half
\Big(
L^a\wedge L_a +\frac{1}{4}L^\a\wedge L_\a
+\frac{1}{4}\Lt{}{\a}\wedge \Lt{\a}{}
\Big).
\eea
This is identical, up to total derivative terms, to
the conventional Wess-Zumino term,
\bea
b_{NS}=\half
\Big[
dx^a\wedge
\Big((\thetab\ga_ad\theta)-(\thetatb\ga_ad\thetat)\Big)
+\half(\thetatb\ga^ad\thetat)\wedge (\thetab\ga_ad\theta)
\Big].
\label{bonF}
\eea
We use the capital letter $B$ as the supersymmetric form of
the Wess-Zumino
term $b$
to avoid confusion.

In turn, the D-string superalgebra is generated by
supertranslations and D-string charges
$T_A = \{ Q_A, \S^A \}$, where
$\S^A=(\S^a,\S^\a,\St^\a)$, as
\bea
\{Q_\a,Q_\b\} &=&
   (\ga^a\otimes I)_{\a\b}P_a
   +(\ga_a\otimes \sigma_1)_{\a\b}\S^a,\nonum\\
\[P_a,Q_\b\]&=&2(\ga_a\otimes I)_{\a\b}\S^\a,\\
\[\S^a,Q_\b\]&=&2(\ga^a\otimes \sigma_1)_{\a\b}\S^\a.\nonum
\eea
Parametrizing the supergroup manifold by
\bea
U=e^{y_a\S^a}e^{\phit_\a\St^\a}e^{\phi_\a\S^\a}
e^{x^aP_a}e^{\thetat^\a \Qt_\a}e^{\theta^\a Q_\a},
\eea
the pullback vielbeins of the left-invariant supergroup
were obtained.
As before, these are superinvariant.
The pullback to the D-string worldsheet
of the R$\otimes$R two-form gauge potential
was written down, in terms of supercurrents, as
\bea
B_R=-\frac{1}{4}
\Big(
L^\a\wedge {\Lh{\a}{}}-\Lt{}{\a}\wedge {\Lth{\a}{}}
\Big),
\label{bhonDL}
\eea
where a caret on $L$ represents a left-invariant pullback vielbein
corresponding to D-string generators $\S^A$.
This was found to be the well-known Wess-Zumino term for D-strings,
up to total derivative terms,
\bea
b_R =\half\Big[
dx^a\wedge
\Big((\thetab\ga_ad\theta)-(\thetatb\ga_ad\thetat)\Big)
+\half(\thetatb\ga^ad\thetat)\wedge (\thetab\ga_ad\theta)
\Big],
\label{bhonD}
\eea
which is identical to the right hand side (R.H.S.) of (\ref{bonF}).
\sect{The NS$\otimes$NS and The R$\otimes$R
Gauge Potentials on The  Worldsheet}
In order to describe $(p,q)$-strings,
we need
the pullback to the {\it same} worldsheet
of the NS$\otimes$NS and the R$\otimes$R two-form gauge potentials,
$b_{NS}$ and $b_{R}$.
In this section, we obtain the pullback to the F-string
worldsheet of the R$\otimes$R two-form gauge potential
(or, the pullback to the D-string worldsheet of the NS$\otimes$NS
gauge potential)
in the following two ways:
First, we consider a map which
transforms the D-string superalgebra
to the F-string superalgebra.
Second, we modify the D-string superalgebra,
and show that the two-form potential can be constructed
by the supercurrents
on the corresponding supermanifold.

We denote the pullback to the D-string worldsheet
of the R$\otimes$R gauge potential as `$b_R$
on the D-string worldsheet $\CW_D$'
and the pullback to the F-string worldsheet
of the R$\otimes$R gauge potential as `$b_R$
on the F-string worldsheet $\CW_F$',
for simplicity.
\subsection{Mapping The D-string Superalgebra
to The F-string Superalgebra}
Let us consider
a map which transforms the D-string superalgebra
to the F-string superalgebra.
The dual map transforms the $b_R$ on the D-string supergroup manifold
$\CM_D$
to a two-form on the F-string supergroup manifold $\CM_F$.
We regard the pullback of the obtained two-form as
the two-form on the F-string worldsheet,
since we have regarded the pullback of the $b_R$
on the D-string supergroup manifold as the $b_R$
on the D-string worldsheet in the previous section.
In this sense,
the obtained two-form can be regarded as
the $b_R$ on the F-string worldsheet $\CW_F$.
There is another point of view,
which regards the two-form on the F-string supergroup manifold
as $b_{NS}$.
The pullback map thus transforms $b_R$ to $b_{NS}$
on the D-string worldsheet
$\CW_D$.
Here, we adopt the former viewpoint below.

The map is summarized as
\bea
\begin{tabular}{cccccc}
$b_R$ on $\CW_D$ &
$\CW_D$&
$\hookrightarrow$&
$\CM_D$
&
$\leftrightarrow$&
$T_A=\{Q_A,\S^A\}$
\\
$\downarrow\hs{-2}\ast$&
$\uparrow$&
&
$\uparrow$&
&
$\downarrow$
\\
$b_R$ on $\CW_F$&
$\CW_F$&
$\hookrightarrow$&
$\CM_F$ 
&
$\leftrightarrow$&
$T_A'=\{{Q_A}', {Z^A}'\}$,
\end{tabular}
\eea
where $\CM_D\ni (x^a, \theta^\a, \thetat^\a, y_a, \phi_\a, \phit_\a)$ and
$\CM_F\ni (x^a, \theta^\a, \thetat^\a, z_a, \zeta_\a, \zetat_\a)$
are the D-string 
and F-string supergroup manifolds, respectively.
The $\ast$ means the pullback.
From the above diagram,
one finds that a map: $b_R$ on $\CW_D$ $\to{\ast}$ $b_R$ on $\CW_F$,
is induced by a map: ${T_A}\rightarrow{T_A}'$.
Thus, we first consider the map
which transforms the generators of D-string superalgebra
to those of F-string superalgebra.
We find that the D-string superalgebra is mapped to
the F-string superalgebra
by
\bea
&&{P_a}'=P_a,~~~
{Q_\a}'=\frac{\ep}{\sqrt{2}}(Q+\Qt),~~~
\Qt_\a{}'=\frac{\ep}{\sqrt{2}}(-Q+\Qt),\nonum\\
&&{Z^a}'=\S^a,~~~
{Z^\a}'=\frac{\ep}{\sqrt{2}}(\S^\a+\St^\a),~~~
{{{\Zt}}^\a}{}'=\frac{\ep}{\sqrt{2}}(\S^\a-\St^\a),
\label{T'=T}
\eea
where $\ep^2=1$.
This is a part of the $SO(2)$ R-symmetry.
We next consider a map, which transforms
spacetime coordinates on $\CM_F$ to those on $\CM_D$.
The corresponding map of spacetime coordinates,
associated with the above map of generators,
is defined by
\bea
e^{{z_a}'{Z^a}'}e^{{\zetat_\a}'{\Zt^\a}{}'}e^{{\zeta_\a}'{Z^\a}'}
e^{{x^a}'{P_a}'}e^{{\thetat^\a}{}'{\Qt_\a}'}e^{{\theta^\a}'{Q_\a}'}
=e^{y_a\S^a}e^{\phit_\a\St^\a}e^{{\phi_\a}{\S^\a}}
e^{{x^a}{P_a}}e^{{\thetat^\a}{\Qt_\a}}e^{{\theta^\a}{Q_\a}}.
\label{e=e}
\eea
After some algebra,
we obtain the following transformations:
\bea
x^a={x^a}',~~~
\theta=\frac{\ep}{\sqrt{2}}(\theta'-\thetat'),~~~
\thetat=\frac{\ep}{\sqrt{2}}(\theta'+\thetat'),~~~
y_a={z_a}'+\half(\thetatb'\ga_a\theta'),
\eea
where expressions for $\phi_\a$ and $\phit_\a$
are omitted.
Performing the above transformations
on $b_R$ on $\CW_D$, (\ref{bhonD}),
we obtain
$b_R$ on $\CW_F$
(or, $b_{NS}$ on $\CW_D$)
as (dropping primes)
\bea
b_R=-\half\Big[
dx^a\wedge \Big( (\thetab\ga_ad\thetat)+(\thetatb\ga_ad\theta)\Big)
+\frac{1}{4}\Big( (\thetab\ga^ad\theta)+(\thetatb\ga^ad\thetat) \Big)
\wedge
\Big( (\thetatb\ga_ad\theta)+(\thetab\ga_ad\thetat) \Big)
\Big].
\label{bhonF}
\eea

We now write two-form $b_R$ on $\CW_F$
in a manifestly supersymmetric form.
To do this, we note
that the left-invariant Maurer-Cartan one-form
of the F-string and D-string superalgebra satisfies
\bea
L^AT_A={L^A}'{T_A}',
\eea
because of (\ref{e=e}).
This implies the following relations:
\bea
&&L^a={L^a}',~~~
L^\a= \frac{\ep}{\sqrt{2}}({L^\a}'-{\Lt{}{\a}}{}'),~~~
\Lt{}{\a}=\frac{\ep}{\sqrt{2}}({L^\a}'+{\Lt{}{\a}}{}'),\nonum\\
&&\hat L_a={L_a}',~~~
\hat L_\a=\frac{\ep}{\sqrt{2}}({L_\a}'+\Lt{}{\a}{}'),~~~
\Lth{\a}{}=\frac{\ep}{\sqrt{2}}({L_\a}'-{\Lt{\a}{}}{}'),
\eea
where a caret represents left-invariant vielbeins
corresponding to D-string charges.
Using the above relations in the supersymmetric form
$\hat B$, (\ref{bhonDL}), of $b_R$ on $\CW_D$,
we obtain the supersymmetric form of $b_R$ on $\CW_F$ as
\bea
B_R=\frac{1}{4}({L^\a}'\wedge {\Lt{\a}{}}{}'
                  -{\Lt{}{\a}}{}'\wedge {L_\a}').
\label{BhonF}
\eea
This is found to be identical to (\ref{bhonF}),
up to total derivative terms,
as expected.
In this way, we have constructed $b_R$ on $\CW_F$ explicitly.
Similarly, $b_{NS}$ on $\CW_D$ is constructed from
$b_{NS}$ on $\CW_F$, in the same manner.
The obtained expression for $b_{NS}$ on $\CW_D$
is identical to the R.H.S of (\ref{bhonF}).
\subsection{Interchanging Fermionic Generators}
We have obtained $b_R$ on $\CW_F$ by a map
which relates the D-string superalgebra
to the F-string superalgebra.
Here, we consider another course
to construct $b_R$ on $\CW_F$.
It was suggested~\cite{sakagu} that
a superalgebra, which is obtained by
interchanging $\S^\a$ and $\St^\a$ in the D-string superalgebra,
describes $b_{NS}$ on $\CW_D$ (or, equivalently $b_{R}$ on $\CW_{F}$).
In fact, using a superalgebra which is obtained by
interchanging fermionic generators,
$\S^\a$ and $\St^\a$, in the D-string superalgebra,
we can construct $B_{NS}$ on $\CW_D$ (or, $B_{R}$ on $\CW_F$),
which is identical to the R.H.S. of (\ref{bhonF})
up to total derivative terms.\footnote{
This turns out to be identical to the two-form found in \cite{sakagu}.
In fact, using a Fierz identity
\bea
(\thetab\ga^ad\theta)\wedge (\thetatb\ga_ad\theta)
=\frac{1}{3}(\thetab\ga^ad\theta)\wedge (\thetab\ga_ad\thetat)
-\frac{1}{3}d\Big((\thetab\ga^ad\theta)(\thetatb\ga_a\theta) \Big),
\nonum
\eea
the last term in (\ref{bhonF}) is rewritten,
up to total derivative terms, as,
$
\frac{1}{3}(\thetab\ga^ad\theta)\wedge (\thetab\ga_ad\thetat)
+\frac{1}{3}(\thetatb\ga^ad\thetat)\wedge (\thetatb\ga_ad\theta).
$
}
This implies that interchanging fermionic generators,
$\S^\a$ and $\St^\a$, corresponds to exchanging the
R$\otimes$R gauge potential for the NS$\otimes$NS one
(or exchanging the D-string worldsheet for the F-string worldsheet).
In this way, we find that the pullback
of the R$\otimes$R and the NS$\otimes$NS
gauge potentials to the same worldsheet
can be described by the F-string superalgebra and a superalgebra
which is obtained by interchanging $\S^\a$ and $\St^\a$ in the
D-string superalgebra.

In order to confirm further the non-trivial feature:
interchanging fermionic generators results in
exchanging the NS$\otimes$NS two-form for
the R$\otimes$R one
(or, interchanging the F- and the D-string worldsheet),
we show that
by starting with a superalgebra, which is obtained by
interchanging fermionic generators, $Z^\a$ and $\Zt^\a$,
in the F-string superalgebra (\ref{Falgebra}),
$b_R$ on $\CW_F$ (or $b_{NS}$ on $\CW_D$)
can be constructed.
The corresponding left-invariant pullback vielbeins
are obtained as
\bea
L^\a&=&d\theta^\a,
   ~~~\tilde L^\a=d\thetat^\a,
   ~~~L^a=dx^a+\half(\thetab\ga^ad\theta)+\half(\thetatb\ga^ad\thetat),\\
L_a&=&dz_a+\half(\thetab\ga_ad\theta)-\half(\thetatb\ga_ad\thetat),\\
\Lt{\a}{}&=&d\zeta_\a
           +2dz_a(\thetab\ga^a)_\a
           +2dx^a(\thetab\ga_a)_\a
           +\frac{2}{3}(\thetab\ga^ad\theta)(\thetab\ga_a)_\a,\\
\L{\a}{}&=&d\zetat_\a
           +2dz_a(\thetatb\ga^a)_\a
           -2dx^a(\thetatb\ga_a)_\a
           -\frac{2}{3}(\thetatb\ga^ad\thetat)(\thetatb\ga_a)_\a.
\eea
We find that the second-order expression,
\bea
-\frac{1}{4}( \L{}{\a}\wedge \L{\a}{}
-\Lt{}{\a}\wedge \Lt{\a}{} ),
\eea
is a supersymmetric form of $b_R$ on $\CW_F$, (\ref{bhonF}), as expected.
In this way, we have seen that the interchanging $Z^\a$ and $\Zt^\a$
results in exchanging the
R$\otimes$R gauge potential for the NS$\otimes$NS one
(or exchanging the D-string worldsheet for the F-string worldsheet).
\sect{The $(p,q)$-String Superalgebra}
In the previous section,
we find that the superalgebra,
\bea
\{Q_\a,Q_\b\}&=&
   (\ga^a\otimes I)_{\a\b}P_a
   +(\ga_a\otimes\sigma_1)_{\a\b}\S^a
   -(\ga_a\otimes\sigma_3)_{\a\b}Z^a,
\label{IIBalgebra1}\\
\[P_a,Q_\b\]&=&
   2(\ga_a\otimes\sigma_1)_{\a\b}\S^\a
   -2(\ga_a\otimes\sigma_3)_{\a\b}Z^\a,\\
\[Z^a,Q_\b\]&=&
   2(\ga^a\otimes I)_{\a\b}Z^\a,\\
\[\S^a,Q_\b\]&=&
   2(\ga^a\otimes I)_{\a\b}\S^\a,
\label{IIBalgebra4}
\eea
describes the pullback to the worldsheet
of the R$\otimes$R and the NS$\otimes$NS two-form
gauge potentials.
Now that we have obtained
the NS$\otimes$NS and the R$\otimes$R gauge potentials
on the same worldsheet,
we consider the $(p,q)$-string superalgebra
which describes the $Sl(2,Z)$-doublet of the two-form gauge potentials.

To do this, we first describe the $Sl(2;Z)$-doublet of
the two-form gauge potentials
in terms of $b_{NS}$ and $b_R$ on the worldsheet, explicitly.
In order to describe the $Sl(2;Z)$-symmetry,
we introduce background scalars which belong to
the coset $Sl(2;Z)/SO(2)$ or, equivalently, $SU(1,1)/U(1)$.
The $Sl(2;Z)$ acts from the left and the $SO(2)$ from the right.
These scalars are represented by a complex $SU(1,1)$
doublet $\cU^r(r=1,2)$.
Using these doublet,
the $SU(1,1)$-invariant complex scalar density is
defined by
\bea
\PsiI=\cU^1\hat\Psi+\cU^2\Psi,
\label{PsiI}
\eea
where $\hat\Psi$ and $\Psi$ form an $Sl(2;Z)$-doublet.
Eliminating an unphysical field by using local $U(1)$ gauge symmetry,
i.e. in the unitary gauge, this is rewritten
as\footnote{
As in \cite{Schwarz},
(\ref{PsiI}) is rewritten, in the unitary gauge,
as $\PsiI=f(\PhiI-B\PhiI^\ast)$,
where $\PhiI$ and $\PhiI^\ast$ are $SU(1,1)$-doublet.
The $f$ is defined by $f=(1-BB^\ast)^{-1/2}$,
and the $B$ transforms linear fractionally
under the finite $SU(1,1)$ group action.
The $B$ has values restricted to the interior
of the unit disc.
The $B$ is expressed in terms of $\tau$
as $B=(\bar\tau+i)/(\bar\tau-i)$,
and $\PhiI$ is in terms of $\hat\Psi$ and $\Psi$ as
$\PhiI=\hat\Psi+i\Psi$.
}
\bea
\PsiI=-ie^{\phi/2}e^{-i\vartheta}(\hat\Psi-\bar\tau\Psi),
\label{PsiIunitary}
\eea
where $e^{i\vartheta}=({\bar\tau-i})/{|\tau+i|}$
and $\tau=\chi+ie^{-\phi}$.
The $\hat\Psi$ and $\Psi$ are linear combinations
of $\Phih$ and $\Phi$
with coefficients written in terms of $\chi$ and $\phi$,
where $\Phih$ and $\Phi$ are
the Hodge duals of the superinvariant
modified two-form field strengths $\hat\CF=d\hat A-b_R$
and $\CF=d A-b_{NS}$, respectively.
These coefficients can be determined by
requiring
that $\PsiI$ should not include $\chi$ and $\phi$ explicitly,
and that $\hat\Psi$ and $\Psi$ should be
$\Phih$ and $\Phi$, respectively,
when $\chi$ and $\phi$ are set to vanish. 
As a result, one finds that
$\hat\Psi$ and $\Psi$ are expressed as
\bea
\hat\Psi&=&
e^{\phi/2}\Re({\tau e^{i\vartheta}})\hat\Phi
-e^{\phi/2}\Im({\tau e^{i\vartheta}})\Phi,
\label{Psih}\\
\Psi&=&
e^{\phi/2}\Re({e^{i\vartheta}})\hat\Phi
-e^{\phi/2}\Im({e^{i\vartheta}})\Phi,
\label{Psi}
\eea
where $\Re(\rho)$ and $\Im(\rho)$ denote the real and imaginary parts
of $\rho$, respectively.
The $\PsiI$ is invariant under the left action of $Sl(2;Z)$:
\bea
\tau\rightarrow \frac{a\tau+b}{c\tau+d},~~~
%
\left( \ba{@{\,}cc@{\,}} a&b \\ c&d\ea \right)\in Sl(2,Z),
\eea
while keeping $\Phih$ and $\Phi$ invariant.
$\hat\Psi$ and $\Psi$ form an $Sl(2;Z)$-doublet up to the right action
of $SO(2)$.

We now consider the $(p,q)$-string superalgebra
which describes the $Sl(2;Z)$-doublet of the two-form
gauge potentials described above.
We first present the $(p,q)$-string superalgebra
and then show that the $Sl(2;Z)$-doublet can be constructed from it.
The $(p,q)$-string superalgebra is generated by $T_A=(Q_A, Z^A, \S^A)$
as
\bea
\{Q_\a,Q_\b\}&=&(\ga^a\otimes I)_{\a\b}P_a
                +(\ga_a\otimes\sigma_1)_{\a\b}(\a\S^a-\ga Z^a)
                +(\ga_a\otimes\sigma_3)_{\a\b}(\b\S^a-\de Z^a),
\label{pqalgebra1}\\
\[P_a,Q_\b\]&=&2(\ga_a\otimes\sigma_1)_{\a\b}(\a\S^\a-\ga Z^\a)
              +2(\ga_a\otimes\sigma_3)_{\a\b}(\b\S^\a-\de Z^\a),\\
\[Z^a,Q_\b\]&=&2(\ga^a\otimes I)_{\a\b}Z^\a,\\
\[\S^a,Q_\b\]&=&2(\ga^a\otimes I)_{\a\b}\S^\a,\label{pqalgebra4}
\eea
where $\a, \b, \ga$ and $\de$ are expressed
in terms of $\chi$ and $\phi$ as
\bea
\left( \ba{@{\,}cc@{\,}} \a & \b \\ \ga &\de \ea \right)=
 e^{\phi/2}\left( \ba{@{\,}cc@{\,}}
\Re(\tau e^{i\vartheta}) & -\Im(\tau e^{i\vartheta})\\
\Re(e^{i\vartheta})& -\Im(e^{i\vartheta})
\ea \right).\label{abcd unitary}
\eea
Note that our superalgebra contains
fermionic charges, $Z^\a$ and $\S^\a$,
as well as bosonic charges, $Z^a$ and $\S^a$,
and these charges transform under the $Sl(2;Z)$.
The left-invariant vielbeins on the corresponding group manifold
are calculated as
\bea
L^\a&=&
    d\theta^\a,~~~~
L^a=
    dx^a
    -\half(\thetab\ga^a\otimes Id\theta),
\\
L_a&=&
    dz_a
    -\half\Big(
    \ga(\thetab\ga_a\otimes \sigma_1d\theta)
    +\de(\thetab\ga_a\otimes\sigma_3d\theta)
    \Big),
\\
\hat L_a&=&
    dy_a
    +\half\Big(
    \a(\thetab\ga_a\otimes\sigma_1d\theta)
    +\b(\thetab\ga_a\otimes\sigma_3d\theta)
    \Big),
\\
L_\a&=&
    2dz_a(\thetab\ga^a\otimes I)_\a
    +d\zeta_\a
    -2dx^a\Big(
    \ga(\thetab\ga_a\otimes\sigma_1)_\a
    +\de(\thetab\ga_a\otimes\sigma_3)_\a
    \Big)
\nonum\\&&
    -\frac{1}{3}(\thetab\ga^a\otimes Id\theta)
    \Big(
    \ga(\thetab\ga_a\otimes\sigma_1)_\a
    +\de(\thetab\ga_a\otimes\sigma_3)_\a
    \Big)
\nonum\\&&
    -\frac{1}{3}
    \Big(
    \ga(\thetab\ga_a\otimes\sigma_1d\theta)
    +\de(\thetab\ga_a\otimes\sigma_3d\theta)
    \Big)
    (\thetab\ga^a\otimes I)_\a,
\\
\hat L_\a&=&
    2dy_a(\thetab\ga^a\otimes I)_\a
    +d\phi_\a
    +2dx^a\Big(
    \a(\thetab\ga_a\otimes\sigma_1)_\a
    +\b(\thetab\ga_a\otimes\sigma_3)_\a
    \Big)
\nonum\\&&
    +\frac{1}{3}(\thetab\ga^a\otimes Id\theta)
    \Big(
    \a(\thetab\ga_a\otimes\sigma_1)_\a
    +\b(\thetab\ga_a\otimes\sigma_3)_\a
    \Big)
\nonum\\&&
    +\frac{1}{3}
    \Big(
    \a(\thetab\ga_a\otimes\sigma_1d\theta)
    +\b(\thetab\ga_a\otimes\sigma_3d\theta)
    \Big)
    (\thetab\ga^a\otimes I)_\a.
\eea
Using these supercurrents,
we find that
the supersymmetric form of the $Sl(2;Z)$-doublet,
$\hat B$
and
$B$,
are constructed as
\bea
\hat B&=&-\half(
\L{}{a}\wedge\Lh{a}{}+\frac{1}{4}\L{}{\a}\wedge\Lh{\a}{}
),\\
B&=&\half(
\L{}{a}\wedge\L{a}{}+\frac{1}{4}\L{}{\a}\wedge\L{\a}{}
).
\eea
In fact, these turn out to be
\bea
\bh&=&\a b_R+\b b_{NS},\\
b&=&\ga b_R+\de b_{NS},
\eea
up to total derivative terms,
where $b_{NS}$ and $b_R$ are defined in (\ref{bonF}) and (\ref{bhonF}),
respectively.
For constant dilaton and axion,
the Hodge duals of the modified field strengths constructed from
the above $Sl(2;Z)$-doublet, $\hat b$ and $b$,
are $\a\Phih +\b \Phi$ and $\ga\Phih+\de \Phi$,
which are nothing but (\ref{Psih}) and (\ref{Psi}), respectively.
We thus regard the superalgebra
(\ref{pqalgebra1})$\sim$(\ref{pqalgebra4})
as the $Sl(2;Z)$-{\it covariant} $(p,q)$-string superalgebra.
At a special point on the $Sl(2;Z)$-orbit, $\chi=\phi=0$,
this superalgebra
reduces to (\ref{IIBalgebra1})$\sim$(\ref{IIBalgebra4}),
which is T-dual to the IIA superalgebra
in presence of F-string, D0 and D2-branes~\cite{sakagu}.

R-symmetry of the superalgebra
is realized as the right action of $SO(2)$.
In fact,
the superalgebra is found to be invariant
(with a trivial scaling of generators
$\S^A$ and $Z^A$) under
R-symmetry transformation:
$Q_\a\rightarrow RQ_\a$, $\S^\a\rightarrow
R\S^\a$ and $Z^\a\rightarrow RZ^\a$,
where $R\in SO(2)$,
associated with the right action of $SO(2)$.
In the case that $\chi=\phi=0$,
the $SO(2)$ gauge transformation
causes a rotation of $\S^A$ and $Z^A$.
Thus, interchanging $\S^A$ and $Z^A$
is a part of R-symmetry.
This is similar to the case encountered in
the map (\ref{T'=T}).
\sect{Summary and Discussions}
We obtained the pullback to the F-string worldsheet
of the R$\otimes$R gauge potential $b_R$
by considering a map,
which transforms the F-string superalgebra to
the D-string superalgebra,
or starting from a superalgebra,
which is obtained by interchanging $\S^\a$ and $\St^\a$
in the D-string superalgebra.
Considering
the $Sl(2;Z)$-doublet of the two-form gauge potentials
explicitly in terms of the obtained $b_R$ and $b_{NS}$,
we presented the $Sl(2,Z)$-{\it covariant}
$(p,q)$-string superalgebra.
It was shown that the $Sl(2;Z)$-doublet
can be constructed by the $(p,q)$-string superalgebra.
The superalgebra at a special point on the $Sl(2;Z)$-orbit,
$\chi=\phi=0$,
reduces to the IIB superalgebra~\cite{sakagu}
which is related by T-duality
to the IIA superalgebra describing F-strings, D0- and D2-branes.

Let us comment on the generalization of our formulation
to the D3-brane, the $(p,q)$ 5-brane and the KK5-brane.
The D3-brane superalgebra
will be obtained as a T-dual
to the IIA superalgebra
in presence of
the D2-brane $\S^{AB}$ and
the D4-brane $\S^{A_1\cdots A_4}$.
In addition,
one finds that the algebraic consistency
requires the F-string $Z^A$ and the D0-brane $\S$.
The D3-brane superalgebra is thus generated
by generators:
not only
the supertranslation $Q_A$ and
the D3-brane $\S^{ABC}$
but also
the F-string $Z_A$
and
the D-string $\S^A$.
This is consistent with the fact
that the field strength $H$ of the four-form gauge potential $C$,
which naturally couples a D3-brane,
is expressed as
$H=dC+\Im({\cal C} \wedge \bar{\cal H})$,
where $\cal C$ represents the SU(1,1)-invariant two-form
gauge potential and $\cal H$ the field strength.
We anticipate that
the supersymmetric form of the four-form gauge potential
$C$ can be constructed
by using the D3-brane superalgebra.
It is a general feature of our superalgebras
that the $p$-brane superalgebra
contains the $p'$-brane superalgebra $(p'< p\leq 5)$
as a subalgebra.
Since the superalgebra, which is T-dual to
the IIA superalgebra in presence of the F-string,
the D0- and the D2-brane,
naturally describes the pullback to the same worldsheet
of the NS$\otimes$NS
and the R$\otimes$R gauge potentials,
the D3-brane superalgebra may describe
the D3-brane on which the $(p,q)$-string sticks.
If this is the case,
it is interesting to study
the effects of interchanging fermionic generators
of the D3-brane charges as well as the $(p,q)$-string charges,
and to examine how the intersecting branes
can be described in our formulation.

In turn, the $(p,q)$ 5-brane superalgebra is generated by generators:
the D5-brane
and
the NS 5-brane
in addition to those of the D3-brane superalgebra.
The 126 bosonic D5-brane charges $\S^{m_1\cdots m_5}$,
$m_i=1,\cdots ,9$,
are related by a T-duality
to the 70 D4-brane charges $\S^{p_1\cdots p_4}$
and the 56 D6-brane charges $\S^{0p_1\cdots p_3}$
as $\S^{\sh p_1\cdots p_4}=\S^{p_1\cdots p_4}$
and $\S^{p_1\cdots p_5}\sim \S^{0p_1\cdots p_3}$, respectively,
where ``$\sim $'' denotes the equivalence modulo the self-duality relation
and $p_i$ runs from $1$ to $9$ except for $\sh$.
Thus, the D5-brane charges are naturally expressed
in terms of the charges $\S^{i_1\cdots i_4}$ in the IIA superalgebra,
or equivalently $Z^{\na i_1\cdots i_4}$ in the M-algebra,
where $\na$ denotes the 11-th direction.
Similarly,
the 126 IIB NS5(KK5)-brane charges $Z_B^{m_1\cdots m_5}$
($Z_B^{0m_1\cdots m_4}$)
are related by a T-duality to
the 70 IIA NS5(KK5)-brane charges $Z^{\sh p_1\cdots p_4}$
($Z^{0p_1\cdots p_4}$)
and the 56 IIA KK5(NS5)-brane charges $Z^{\sh 0p_1\cdots p_3}$
($Z^{p_1\cdots p_5}$)
as
$Z_B^{\sh p_1\cdots p_4} =Z^{\sh p_1\cdots p_4}~~
(Z_B^{0 p_1\cdots p_4} =Z^{0 p_1\cdots p_4})$
and
$Z_B^{p_1\cdots p_5} \sim Z^{\sh 0p_1\cdots p_3}
(Z_B^{0\sh p_1\cdots p_3} \sim Z^{p_1\cdots p_5}
)$,
respectively.
It follows that the IIB NS5(KK5)-brane charges are
expressed in terms of the charges
$Z^{\sh i_1\cdots i_4}~~
(Z^{i_1\cdots i_5})
$
in the IIA superalgebra and the M-algebra.
But it is not clear
how the charges with spinorial indices
are related to the IIA superalgebra.
The D5- and NS5-brane charges with spinorial indices,
except for those with five spinorial indices,
are expected to be determined by
the charges
$Z^{\na A_1\cdots A_4}$
and 
$Z^{\sh A_1\cdots A_4})$
in the M-algebra, respectively.
But the charges with five spinorial indices
will not be related to the charges
of the M-algebra.
If this is the case,
we must add these charges with five fermionic indices
in order to construct the corresponding gauge potentials
in the manifestly supersymmetric forms.
In the case of the KK5-brane charges,
the situation is more intricate
since the T-dual of the corresponding IIA superalgebra
is not rewritten in a covariant form.
Thus,  the charges
constituting the $(p,q)$5-brane and KK5-brane superalgebra
are not determined in terms of
those of the IIA superalgebra completely,
unlike the D3-brane superalgebra.
We hope to report on these issues in the near future~\cite{sakagu2}.
\vs{5}
~\\
{\bf Acknowledgments}

The author would like to express his gratitude to Prof. N. Ishibashi.
This research was supported in part by JSPS Research Fellowships
for Young Scientists.

\end{document}